\documentclass[journal,twocolumn]{IEEEtran}
\IEEEoverridecommandlockouts
\usepackage{balance}
\usepackage{cite}
\usepackage{amsmath,amssymb,amsfonts,amsthm}
\usepackage{graphicx}
\usepackage{graphics}
\usepackage{textcomp}

\newtheoremstyle{case}{}{}{}{}{}{:}{ }{}
\theoremstyle{case}

\usepackage{xcolor}
\usepackage[short]{optidef}
\def\BibTeX{{\rm B\kern-.05em{\sc i\kern-.025em b}\kern-.08em
    T\kern-.1667em\lower.7ex\hbox{E}\kern-.125emX}}
    
\begin{document}

\title{Lightwave Power Transfer-Enabled Underwater Optical ISAC Systems under Ship Attitude Variation\\ \vspace{-3mm}}

\author{
\IEEEauthorblockA{\normalsize{Kapila W. S. Palitharathna}\IEEEauthorrefmark{1}, \normalsize{Constantinos Psomas}\IEEEauthorrefmark{2}, and \normalsize{Ioannis Krikidis}\IEEEauthorrefmark{1}}\\
\IEEEauthorblockA{\IEEEauthorrefmark{1}Department of Electrical and Computer Engineering, University of Cyprus, Nicosia, Cyprus\\}
\IEEEauthorblockA{\IEEEauthorrefmark{2}Department of Computer Science and Engineering, European University Cyprus, Nicosia, Cyprus\\}
\IEEEauthorblockA{Email: \{palitharathna.kapila, krikidis\}@ucy.ac.cy}, c.psomas@euc.ac.cy\vspace{-11mm}}

\maketitle

\begin{abstract}
   In this paper, we propose a lightwave power transfer-enabled underwater optical integrated sensing and communication (O-ISAC) system, where an access point (AP) mounted on a seasurface ship transmits lightwave signals to two nodes, namely ($i$) a seabed sensor that harvests energy and transmits uplink information to the AP, and ($ii$) a sensing target whose position is estimated by the AP using an array of pinhole cameras. To capture practical deployment conditions, the ship attitude variation is modeled through its roll, pitch, and yaw angles, each following a Gaussian distribution under low-to-moderate sea states. Closed-form approximations are derived for the mean squared error (MSE) of target localization and the achievable uplink data rate. Analytical and simulation results demonstrate excellent agreement, validating the proposed models and derived expressions, while revealing the fundamental communication-sensing tradeoff in the O-ISAC system. The results further provide valuable design insights, including the optimal camera placement on the ship to minimize localization error, achieving a minimum MSE of $10^{-2}$ $\text{m}^2$ with multiple cameras under roll, pitch, and yaw angle variation of $10^{\circ}$, and the optimal harvest-use ratio of $0.55$ for the considered setup.
\end{abstract}

\begin{IEEEkeywords}
Underwater optical wireless communication, optical ISAC, lightwave power transfer, ship attitude variation.
\end{IEEEkeywords}

\vspace{-5mm}
\section{Introduction}
With the increasing demand for high-speed data transmission in underwater environments, reliable communication technologies are critical for applications such as ocean exploration, environmental monitoring, offshore industry operations, and defense. Traditional acoustic systems, although widely deployed, are limited by low bandwidth and high latency, rendering them unsuitable for data-intensive applications. Concurrently, underwater optical wireless communication (UOWC) has attracted significant attention due to its capability to provide high-capacity, and low-latency links. In particular, UOWC systems has been proposed as a promising communication medium to collect data from seabed sensors periodically~\cite{Zeng_2017}.  

Despite these advances, UOWC systems face significant challenges due to attenuation, turbulence effect, and transmitter/receiver misalignment, which limit both coverage and reliability~\cite{Zeng_2017}. To combat such effects, various techniques have been proposed, including relay-assisted multi-hop systems~\cite{Celik_2020}, multi-antenna systems with antenna selection strategies~\cite{Kapila_2021}, and aperture averaging lenses~\cite{Xiang_2024}. Most existing studies focus primarily on channel impairments caused by submerged particles and oceanic currents. However, the swaying motion, and attitude variation of sea-surface ships or buoys often serving as UOWC access points (APs) remains largely neglected, despite its inevitable impact on system performance and the need for further investigation~\cite{Newman_1977}.

Underwater communication systems rely on battery-powered submerged sensors/nodes, which inherently limits their operational lifetime and makes battery replacement challenging. Energy harvesting from sources such as solar radiation or radio frequency (RF) signals is often impractical in deep-sea environments due to their unavailability. To address this limitation, lightwave power transfer (LPT) has been proposed as a solution, enabling energy transfer from sea-surface ships/buoys to seabed sensors using lightwave signals, which can then be used for uplink (UL) communication~\cite{Kapila_2022}. In this context, simultaneous lightwave information and power transfer (SLIPT) protocols offer a promising approach~\cite{Panagiotis_2018}. Specifically,~\cite{Shuai_2019} proposes a novel SLIPT architecture that employs a photodiode (PD) as the information receiver and a solar panel as the energy harvester. Furthermore, the results in~\cite{Uysal_2021} demonstrate that the energy harvested via SLIPT is sufficient to recharge an underwater sensor node confirming its feasibility.

On a parallel development, optical integrated sensing and communication (O-ISAC) leverages the large bandwidth of optical signals to enable high-speed communication and high-resolution sensing, offering advantages over RF-based ISAC in free-space optical systems~\cite{Wen_2024}. Research on O-ISAC spans both wired domains (e.g., fiber-optic ISAC) and wireless domains (e.g., free-space optics-based ISAC)~\cite{Wen_2024}. Most existing wireless O-ISAC implementations employ laser sources as transmitters emitting coherent light and PDs as receivers, making them functionally similar to RF-ISAC~\cite{Wen_2024,Alizera_2025}. In contrast, an O-ISAC framework based on cost-effective commercial light-emitting diodes (LEDs) as transmitter and pinhole cameras as receivers has been studied in~\cite{Zhang_2025}. However, no prior work has applied O-ISAC for underwater systems. We identify that O-ISAC is a promising approach for underwater sensing, an area of rapidly growing interest in oceanography. Specifically, the LED/pinhole camera-based O-ISAC systems can mitigate the effects of transmitter/receiver misalignment, turbulence, swaying, attitude variation. To the best of the authors' knowledge, this is the first work to study LPT-enabled underwater O-ISAC and analyze its performance under ship attitude variation. 

In this paper, we propose an LPT-enabled O-ISAC system, where a sea-surface AP mounted on a ship/buoy transmits lightwave signals to seabed sensors. The energy-harvesting (EH) sensor harvests energy from the received optical signals for uplink transmission, while a sensing target reflects the light for localization at the AP using pinhole cameras. Specifically, we consider the practical effect of ship attitude variations caused by oceanic waves, modeled through the roll, pitch, and yaw angles of the ship. To evaluate system performance, we derive closed-form expressions for the sensing mean squared error (MSE) and the average achievable uplink rate under the Gaussian assumption for the ship's roll, pitch, and yaw angles. The obtained results validate the analytical framework and reveal the communication–sensing tradeoff inherent in the proposed O-ISAC system. In particular, they help to identify the optimal time allocation factor between downlink and uplink operations, as well as the optimal placement of the pinhole cameras to enhance system performance.

\begin{figure}
    \centering
    \includegraphics[width=0.6\linewidth]{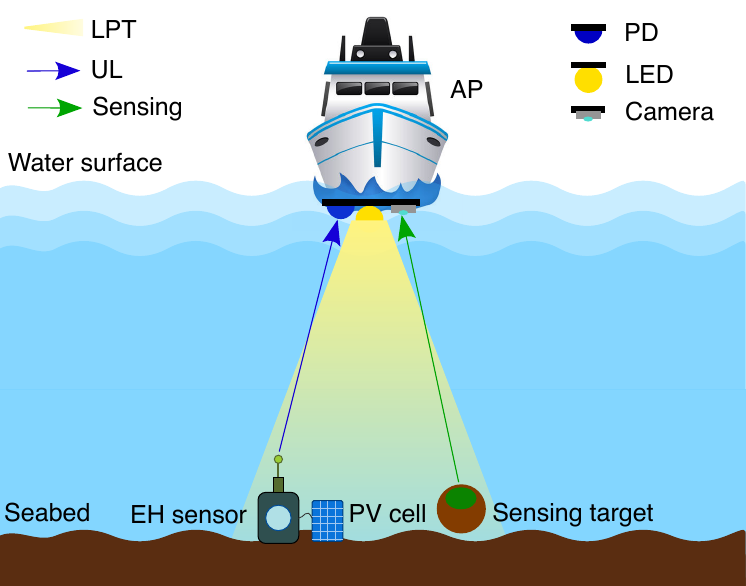}
    \vspace{-2mm}
    \caption{LPT-enabled O-ISAC system consisting of a sea-surface AP as well as an EH sensor, and a sensing target at the seabed.}
    \label{fig:system}
    \vspace{-7mm}
\end{figure}

\vspace{-5mm}
\section{System model}
We consider an LPT-enabled O-ISAC system consisting of a sea-surface AP, an EH sensor, and a sensing target, as illustrated in Fig.~\ref{fig:system}. The sea-surface AP is mounted on a ship/buoy and is connected to a fixed power supply, whereas the EH sensor lacks a dedicated power source and must harvest optical energy to enable its uplink communication with the AP. Specifically, the EH sensor captures the lightwave energy transmitted by the LED attached to the AP in the downlink through a photovoltaic (PV) cell, and subsequently uses the harvested energy to transmit its backlogged data through its own LED to the PD at the AP during the uplink phase. To incorporate LPT, the system operates in frames of duration $T$. During the initial $\alpha T$ portion of each frame, where $0<\alpha\le 1$, downlink LPT and sensing are performed, while uplink communication occurs during the remaining $(1-\alpha)T$ interval. For simplicity, we assume that the LED on the EH sensor and the PD on the AP are well aligned~\cite{Kapila_2022}. Meanwhile, the sensing target reflects a portion of the received optical signal back towards the AP for localization. The ship/buoy is equipped with $M$ pinhole cameras mounted on a flat surface to capture the reflected light from the sensing target and estimate its position. The 3D locations of the AP, sensing target, and EH sensor in the global coordinate system are denoted by $\mathbf{p}_{A} = [x_{A},\, y_{A},\, z_{A}]^{\mathsf{T}}$, 
$\mathbf{p}_{S} = [x_{S},\, y_{S},\, z_{S}]^{\mathsf{T}}$, and 
$\mathbf{p}_{E} = [x_{E},\, y_{E},\, z_{E}]^{\mathsf{T}}$, respectively.

In this work, we consider the practical effect of ship attitude variations due to oceanic waves, which have been largely neglected in existing UOWC literature. Specifically, we model the sea-surface as a stationary Gaussian random process, and the ship's response to sea-surface variations as a linear system~\cite{Newman_1977}. Under these assumptions, the roll, pitch, and yaw angles of the ship, denoted by $\theta_{\text{R}}$, $\theta_{\text{P}}$, and $\theta_{\text{Y}}$, respectively, follow independent Gaussian distributions with probability density function (pdf) 
$f_{\theta_K}(x) = \frac{1}{\sqrt{2\pi}\sigma_{\theta_K}} \exp\!\Big(-\frac{(x - \mu_{\theta_K})^2}{2\sigma_{\theta_K}^2}\Big)$, 
where $K \in \{\text{R}, \text{P}, \text{Y}\}$. The mean values are given by $\mu_{\theta_{\text{R}}} = 0^{\circ}$, $\mu_{\theta_{\text{P}}} = 0^{\circ}$, and $\mu_{\theta_{\text{Y}}} = \theta_D$, while the corresponding variances are $\sigma_{\theta_{\text{R}}}^2$, $\sigma_{\theta_{\text{P}}}^2$, and $\sigma_{\theta_{\text{Y}}}^2$. 
We further assume that the PD, LED, and pinhole cameras are rigidly mounted on the ship (or buoy) and thus experience identical orientation variations.

\vspace{-3mm}
\subsection{Optical Sensing}
We employ the pinhole camera model to project the reflected optical signals onto an image plane for sensing~\cite{Zhang_2025}. Let the 3D position of the target device in the $m$-th camera coordinate frame be denoted as 
$\mathbf{p}_{c,m} = [x_{c,m},\, y_{c,m},\, z_{c,m}]^{\mathsf{T}}$. 
We assume that the spatial relationship among multiple cameras is fixed and can be expressed as
$x_{c,m} = x_{c,1} + \Delta x_m$, $y_{c,m} = y_{c,1} + \Delta y_m$, $z_{c,m} = z_{c,1}$, where $\Delta x_m$ and $\Delta y_m$ are constant offsets determined by the relative geometry of the camera array. 

The coordinates of the target device in the $m$-th camera coordinate frame, $\mathbf{p}_{c,m}$, and in the real-world coordinate frame, $\mathbf{p}_D$, can be transformed into each other as
\vspace{-2mm}
\begin{equation}
    \mathbf{p}_{c,m} = \mathbf{Q}_m^{\mathsf{T}} (\mathbf{p}_S - \mathbf{t}_m),
\end{equation}
where $\mathbf{Q}_m \in \mathbb{R}^{3\times3}$ is the rotation matrix of the $m$-th camera with respect to the real-world frame, and $\mathbf{t}_m \in \mathbb{R}^{3\times1}$ is the translation vector. The sets $\{\mathbf{Q}_m\}$ and $\{\mathbf{t}_m\}$ together form the exterior orientation parameters of the camera array. The rotation matrix $\mathbf{Q}_m$ can be decomposed using roll, pitch, and yaw rotations as \(\mathbf{Q}_m = \mathbf{R}_z(\theta_{\text{Y}})\mathbf{R}_y(\theta_{\text{P}})\mathbf{R}_x(\theta_\text{R})\),
where $\mathbf{R}_x(\theta_{\text{R}})$, $\mathbf{R}_y(\theta_{\text{P}})$, and $\mathbf{R}_z(\theta_{\text{Y}})$ represent the roll, pitch, and yaw rotation matrices, respectively. Expanding these rotations, $\mathbf{Q}_m$ can be written explicitly as
\vspace{-1mm}
\begin{equation}\label{equ:Q_mat}
    \mathbf{Q}_m \hspace{-0.5mm}=\hspace{-1.5mm}
    \setlength{\arraycolsep}{3pt} 
    \begin{bmatrix}
    \text{c}\theta_{\mathrm{Y}}\text{c}\theta_{\mathrm{P}} & 
    \text{c}\theta_{\mathrm{Y}}\text{s}\theta_{\mathrm{P}}\text{s}\theta_{\mathrm{R}} \hspace{-0.5mm}-\hspace{-0.5mm} \text{s}\theta_{\mathrm{Y}}\text{c}\theta_{\mathrm{R}} & 
    \text{c}\theta_{\mathrm{Y}}\text{s}\theta_{\mathrm{P}}\text{c}\theta_{\mathrm{R}} \hspace{-0.5mm}+\hspace{-0.5mm} \text{s}\theta_{\mathrm{Y}}\text{s}\theta_{\mathrm{R}} \\
    \text{s}\theta_{\mathrm{Y}}\text{c}\theta_{\mathrm{P}} & 
    \text{s}\theta_{\mathrm{Y}}\text{s}\theta_{\mathrm{P}}\text{s}\theta_{\mathrm{R}} \hspace{-0.5mm}+\hspace{-0.5mm} \text{c}\theta_{\mathrm{Y}}\text{c}\theta_{\mathrm{R}} & 
    \text{s}\theta_{\mathrm{Y}}\text{s}\theta_{\mathrm{P}}\text{c}\theta_{\mathrm{R}} \hspace{-0.5mm}-\hspace{-0.5mm} \text{c}\theta_{\mathrm{Y}}\text{s}\theta_{\mathrm{R}} \\
    -\text{s}\theta_{\mathrm{P}} & 
    \text{c}\theta_{\mathrm{P}}\text{s}\theta_{\mathrm{R}} & 
    \text{c}\theta_{\mathrm{P}}\text{c}\theta_{\mathrm{R}}
\end{bmatrix}\hspace{-1mm},
\end{equation}
where $\text{s}\theta = \sin(\theta)$ and $\text{c}\theta = \cos(\theta)$. Finally, the translation vector is set to \(\mathbf{t}_m = \mathbf{p}_A\).

Each film has a 2D plane coordinate system. We denote $\mathbf{p}_m = (x_m, y_m)$ as the coordinates of the target on the 2D plane. According to the pinhole imaging principle, $\mathbf{p}_m$ can be obtained from $\mathbf{p}_{c,m}$, and their relationships can be written as 
\vspace{-2mm}
\begin{align}\label{coor_relation}
z_{c,m}[\mathbf{p}_m;1] =\mathbf{K}_m\mathbf{p}_{c,m},    
\end{align}
where \(\mathbf{K}_m = [f_{x,m}, 0, 0; 0, f_{y,m}, 0; 0, 0, 1]\). Here $f_{x,m}$ and $f_{y,m}$ are the focal lengths of the $m$-th pinhole camera, and $\mathbf{K}_m$ denotes the interior orientation parameters (IOPs). We assume all cameras share the same IOPs i.e., $\mathbf{K}_m = \mathbf{K}$ for all $m\in\{1,\ldots,M\}$.

We perform image processing algorithms to estimate the target coordinates, where the  estimation accuracy depends on the light intensity and contrast ratio. The received light intensity is contaminated by additive white Gaussian noise (AWGN), and the coordinate estimation error follows Gaussian distribution when the pixel size is sufficiently small. Hence, the estimated coordinates can be expressed as
\vspace{-2mm}
\begin{align}\label{est_coor}
    \hat{\mathbf{p}}_m = \mathbf{p}_m + \mathbf{e}_m = 
    \left[\begin{array}{cc}
         x_m\\
         y_m 
    \end{array}\right] + \left[\begin{array}{cc}
         e_{x,m} \\
         e_{y,m}
    \end{array}\right],
\end{align}
where the variance of the estimation error is inversely proportional to the light intensity i.e., $e_{x,m}, e_{y,m} \sim \mathcal{N}(0, \eta\sigma_I^2/\alpha I^{ref}_m)$~\cite{Zhang_2025}. Here $\eta$ is a scaling factor determined by the film-plane size and its distance to the pinhole, $\sigma_I^2$ is the AWGN variance in the received light, and $I^{ref}_m$ is the reflected light intensity at the $m$-th camera, given by
\vspace{-2mm}
    \begin{align}
        I^{ref}_m = \frac{\rho_{s}h_{S,A,m}h_{A,S}P_{DL}}{A_{cam}},
    \end{align}
where $A_{cam}$ is the effective area of the pinhole camera, $h_{A,S}$ is the channel gain from the AP to the sensing target, $h_{S,A,m}$ is the channel gain from the sensing target to the $m$-th camera, and $P_{DL}$ is the AP transmit power. From~\eqref{coor_relation} and~\eqref{est_coor}, we have
\vspace{-6mm}
\begin{align}\label{coor_relation2}
    z_{c,m}\left[
         \hat{\mathbf{p}}_m;
         1 \right] = \mathbf{K}\mathbf{p}_{c,m} + z_{c,m}\left[
         \mathbf{e}_m ;
         0 \right].
\end{align}
We assume that all pinhole cameras are mounted on the same plane. After some algebraic manipulations,~\eqref{coor_relation2} can be rewritten in the compact form as
\vspace{-3mm}
\begin{align}
    \boldsymbol{\gamma}
    =
    \mathbf{\Sigma} \,
    \mathbf{p}_{\mathrm{c},1},
\end{align}
where $\boldsymbol{\gamma}=[0, 0, \ldots, -f_x\Delta x_m, -f_y\Delta y_m, \ldots]^T$, and $\boldsymbol{\Sigma} = [f_x, 0, -x_1; 0, f_y, -y_1; \ldots; f_x, 0, -x_m; 0, f_y, -y_m; \ldots]$. Here, $\boldsymbol{\gamma}$ is a constant vector determined by the geometric configuration of the camera array. The estimated matrix $\hat{\boldsymbol{\Sigma}}$, obtained from the estimated pixel coordinates $\hat{\mathbf{p}}_m$ can be expressed as 
\begin{align}\label{sigma1}
    \hat{\boldsymbol{\Sigma}} = \left[\begin{array}{ccc}
         f_x & 0 & -\hat{x}_1\\
         0 & f_y & -\hat{y}_1\\
         \vdots & \vdots & \vdots  \\
         f_x & 0 & -\hat{x}_m \\
         0 & f_y & -\hat{y}_m \\
         \vdots & \vdots & \vdots  \\
    \end{array}\right],
    \vspace{-2mm}
\end{align}
where $(\hat{x}_m,\hat{y}_m)$ denote the estimated 2D coordinates of the sensing target in the $m$-th camera coordinate system. Next, the estimated position vector $\hat{\mathbf{p}}_{C,1}$ is obtained using the least squares method as 
\vspace{-3mm}
\begin{align}
    \hat{\mathbf{p}}_{c,1} = (\hat{\boldsymbol{\Sigma}}^{\mathsf{T}}\hat{\boldsymbol{\Sigma}})^{-1}\hat{\boldsymbol{\Sigma}}^{\mathsf{T}}\boldsymbol{\gamma}.
\end{align}
The corresponding 3D coordinates of the target sensor in the real-world coordinate system are then computed as 
\begin{align}
    \left[\begin{array}{cc}
         \hat{\mathbf{p}}_S \\
         1 
    \end{array}\right] = \left[\begin{array}{cc}
         \mathbf{Q}_1 & \mathbf{t}_1\\
         0 & 1
    \end{array}\right]
    \left[\begin{array}{c}
         \hat{\mathbf{p}}_{c,1}\\
         1
    \end{array}\right].
\end{align}
To evaluate the localization performance, the MSE of the estimated position is defined as
\begin{align}\label{equ:mse_1}
    \text{MSE}_p = \mathbb{E}\left\{||\hat{\mathbf{p}}_S - \mathbf{p}_S||^2\right\}.
\end{align}

\vspace{-4mm}
\subsection{EH and Communication}
In the initial $\alpha T$ duration, LPT occurs from the AP to the EH sensor. The harvested energy at the PV cell of the EH sensor is expressed as~\cite{Panagiotis_2018}
\vspace{-2mm}
    \begin{align}\label{equ:EH1}
        E_h = fv_t\alpha Tr_{PV}h_{A,E}P_{DL}\ln\left(1+\frac{r_{PV}h_{A,E}P_{DL}}{I_0}\right),
    \end{align}
where $f$ is the fill factor, $v_t$ is the thermal voltage, $h_{A,E}$ is the channel gain between the AP and the EH sensor, $r_{PV}$ is the responsivity of the PV cell, and $I_0$ is the dark saturation current. During the remaining $(1-\alpha)T$ time fraction, the harvested energy is utilized for uplink data transmission from the EH sensor to the AP. The uplink transmit power is therefore given by \(P_{UL} = E_H/(1-\alpha)T\). Since the exact capacity of optical wireless channels is unknown, the instantaneous achievable rate lower bound derived in~\cite{Lapidoth_2009} is employed. Accordingly, the achievable uplink rate can be expressed as
    \begin{align}\label{equ:inst1}
        R_{\mathrm{UL}}
        = \frac{1-\alpha}{2} \log_{2} \left(
        1 + \frac{e}{2\pi} \cdot
        \frac{(rh_{E,A} P_{UL})^2}{\sigma_n^2}
        \right),
    \end{align}
where $h_{E,A}$ is the channel gain from the EH sensor to the AP, $r$ is the responsivity of the PD, $e$ is the Euler's number, and $\sigma_n^2$ denotes the variance of the zero mean AWGN.

\vspace{-4mm}
\subsection{Channel Model}
UOWC channels are significantly influenced by attenuation, scattering, and turbulence effects. The overall channel gain between an LED transmitter and a PD receiver, denoted by $h$, can be modeled as the product of three independent factors given by \(h = G_d h_t = h_g h_p h_t\), where $h_g$, $h_p$, and $h_t$ represent the geometric loss, path loss, and turbulence-induced fading, respectively~\cite{Kapila_2021}. $G_d = h_g h_p$ is the deterministic component.

The geometric loss $h_g$ for a line-of-sight (LoS) link is expressed as~\cite{Kapila_2021}
\vspace{-2mm}
\begin{align}\label{equ:GL}
    \hspace{-3mm}h_g =
    \begin{cases}
        \dfrac{(m_1+1)A_p}{2\pi d^2}\cos^m(\theta)\cos(\phi)T c_1(\phi), & \hspace{-3mm}|\theta| \le \frac{\pi}{2},\\
        0, & \hspace{-3mm}\text{otherwise},
    \end{cases}
    \vspace{-4mm}
\end{align}
where $m_1=-\ln(2)/\ln(\cos(\theta_{1/2}))$ is the Lambertian order, $A_p$ is the PD aperture area, $d$ is the transmitter–receiver distance, $\theta$ and $\phi$ denote the irradiance and incidence angles, $T$ is the trans-impedance amplifier gain, and $c(\phi)$ is the concentrator gain given by
\vspace{-2mm}
\begin{align}
    c_1(\phi) =
    \begin{cases}
        \dfrac{\rho^2}{\sin^2(\Phi_{\text{FoV}})}, & 0 \le \phi \le \Phi_{\text{FoV}},\\
        0, & \text{otherwise},
    \end{cases}
\end{align}
with $\rho$ as the refractive index and $\Phi_{\text{FoV}}$ the field-of-view of the concentrator. Light propagation in water experiences absorption and scattering, modeled by Beer's law as~\cite{Kapila_2021} \(h_p = \exp[-c(\lambda)d]\),
where $c(\lambda) = a(\lambda) + b(\lambda)$ is the attenuation coefficient, and $a(\lambda)$ and $b(\lambda)$ denote the absorption and scattering coefficients, respectively. Under weak turbulence, the fading coefficient $h_t$ follows a log-normal distribution, whose pdf is given by~\cite{Kapila_2021}
\vspace{-3mm}
\begin{align}
    f_{h_t}(h_t) = \frac{1}{2h_t\sqrt{2\pi\sigma_x^2}}
    \exp\!\left[-\frac{(\ln(h_t)-2\mu_x)^2}{8\sigma_x^2}\right],
\end{align}
where $\mu_x = -\sigma_x^2$ ensures $\mathbb{E}\{h_t\}=1$, and $\sigma_x^2 = \tfrac{1}{4}\ln(\sigma_I^2 + 1)$ relates to the scintillation index $\sigma_I^2 < 1$.

\vspace{-2mm}
\section{Analysis of the MSE of the Sensing}
In order to derive mathematical expressions for the MSE of the sensing, the following steps are followed. First, $\hat{\boldsymbol{\Sigma}}$ can be simplified using the relation in~\eqref{coor_relation2} and expressed as
\vspace{-2mm}
\begin{align}
    \hat{\boldsymbol{\Sigma}} = \left[\begin{array}{ccc}
         f_x & 0 & -\frac{f_{x}}{z_{c,1}}x_{c,1}-e_{x,1}\\
         0 & f_y & -\frac{f_{y}}{z_{c,1}}y_{c,1}-e_{y,1}\\
         \vdots & \vdots & \vdots  \\
         f_x & 0 & -\frac{f_{x}}{z_{c,1}}x_{c,m}-e_{x,m} \\
         0 & f_y & -\frac{f_{y}}{z_{c,1}}y_{c,m}-e_{y,m} \\
         \vdots & \vdots & \vdots  \\
    \end{array}\right].
\end{align}
The pseudo-inverse $(\hat{\boldsymbol{\Sigma}}^T\hat{\boldsymbol{\Sigma}})^{-1}\hat{\boldsymbol{\Sigma}}^T$ can be computed with straightforward algebraic manipulations and is given in~\eqref{pseudo2}, shown at the top of the next page. Here, $\Delta_1 = ABE-AD^2-BC^2$, $A = M f_x^2$, $B = M f_y^2$, $C = f_x \sum_{i=1}^M \tilde{\alpha}_i$, $D = f_y \sum_{i=1}^M \tilde{\beta}_i$, $E = \sum_{i=1}^M \left( \tilde{\alpha}_i^2 + \tilde{\beta}_i^2 \right)$, $\tilde{\alpha}_i = f_x x_{c,i}/z_{c,1} + e_{x,i}$, and $\tilde{\beta}_i = f_y y_{c,i}/z_{c,1} + e_{y,i}$. 
 \begin{figure*}[!t]
     \begin{align}\label{pseudo2}
         (\hat{\boldsymbol{\Sigma}}^T\hat{\boldsymbol{\Sigma}})^{-1}\hat{\boldsymbol{\Sigma}}^T \hspace{-1mm}=\hspace{-1mm} \frac{1}{\Delta_1}\hspace{-1mm}\Bigg[\hspace{-1.5mm}
         \begin{array}{ccccc}
              f_x (B E \hspace{-0.5mm}-\hspace{-0.5mm} D^2) \hspace{-0.5mm}-\hspace{-0.5mm} B C \tilde{\alpha}_1 & C D f_y - B C \tilde{\beta}_1 & \ldots & f_x (B E \hspace{-0.5mm}-\hspace{-0.5mm} D^2) \hspace{-0.5mm}-\hspace{-0.5mm} B C \tilde{\alpha}_M & C D f_y - B C \tilde{\beta}_M\\
              C D f_x - A D \tilde{\alpha}_1 & f_y (A E \hspace{-0.5mm}-\hspace{-0.5mm} C^2) \hspace{-0.5mm}-\hspace{-0.5mm} A D \tilde{\beta}_1 & \dots & C D f_x - A D \tilde{\alpha}_M & f_y (A E \hspace{-0.5mm}-\hspace{-0.5mm} C^2) \hspace{-0.5mm}-\hspace{-0.5mm} A D \tilde{\beta}_M \\
              B C f_x - A B \tilde{\alpha}_1  & A D f_y - A B \tilde{\beta}_1 & \ldots & B C f_x - A B \tilde{\alpha}_M  & A D f_y - A B \tilde{\beta}_M
         \end{array}\hspace{-2mm}\Bigg],
     \end{align}
     \vspace{-7mm}
     \begin{align}\label{P_C1_1}
        \hat{\mathbf{p}}_{c,1}=-\frac{1}{\Delta_1}\sum_{i=2}^M\Bigg[
        & f_x \Delta x_m
        \begin{bmatrix}
        f_x (B E - D^2) - B C \tilde{\alpha}_i \\
        C D f_x - A D \tilde{\alpha}_i \\
        B C f_x - A B \tilde{\alpha}_i
        \end{bmatrix}
        +
        f_y \Delta y_m
        \begin{bmatrix}
        C D f_y - B C \tilde{\beta}_i \\
        f_y (A E - C^2) - A D \tilde{\beta}_i \\
        A D f_y - A B \tilde{\beta}_i
        \end{bmatrix}
        \Bigg],
    \end{align}
    \hrule
    \vspace{-5mm}
 \end{figure*}
Next, multiplying~\eqref{pseudo2} by $\boldsymbol{\gamma}$ and performing some simplifications, the estimate $\hat{\mathbf{p}}_{c,1}$ can be given in \eqref{P_C1_1}, shown at the top of the page.

The term $\Delta_1$ can be further simplified. Substituting the definitions of $A$, $B$, $C$, $D$, and $E$, using the binomial expansion, and with simple mathematical manipulations, $\Delta_1$ can be expressed as
\vspace{-4mm}
\begin{align}
    \hspace{-3mm}\Delta_1 \hspace{-1mm}=\hspace{-1mm} Mf_x^2f_y^2\bigg[(M\hspace{-1mm}-\hspace{-1mm}1)\hspace{-1mm}\sum_{i=1}^M(\hat{\alpha}_i^2\hspace{-0.5mm}+\hspace{-0.5mm}\hat{\beta}_i^2)\hspace{-0.5mm}-\hspace{-0.5mm}2\hspace{-4mm}\sum_{1\le i<j\le M}\hspace{-4mm}(\hat{\alpha}_i\hat{\alpha}_j\hspace{-0.5mm}+\hspace{-0.5mm}\hat{\beta}_i\hat{\beta}_j)\bigg].\hspace{-2mm}\vspace{-2mm}
\end{align}
We now assume that the error terms $e_{x,m}$ and $e_{y,m}$ are zero mean and uncorrelated, and mutually independent. Under these assumptions, cross error and linear error terms vanish in expectation, and the simplified approximation for $\Delta_1$ becomes
\vspace{-5mm}
\begin{align}\label{equ:MSE_avg}
\hspace{-2mm}
    \Delta_1 &\approx Mf_x^2f_y^2\bigg[(M-1)\sum_{i=1}^M\bigg(\frac{f_x^2x_{c,i}^2}{z_{c,1}^2}+\frac{f_y^2y_{c,i}^2}{z_{c,1}^2}+\mathbb{E}\{\sigma_{x,i}^2\}\nonumber\\
    &+\mathbb{E}\{\sigma_{y,i}^2\}\bigg)
    \hspace{-0.5mm}-\hspace{-0.5mm}2\sum_{i<j}\bigg(\frac{f_x^2x_{c,i}x_{c,j}}{z_{c,1}^2}+\frac{f_y^2y_{c,i}y_{c,j}}{z_{c,1}^2}\bigg)\bigg].
\end{align}
\vspace{-2mm}

Since the rotation from the camera frame to the global frame is orthonormal, the MSE in the world coordinates in~\eqref{equ:mse_1} is given by $\overline{\text{MSE}} = \mathbb{E}\left\{||\hat{\mathbf{p}}_{c,1} - \mathbf{p}_{c,1}||^2\right\}$.
Now, we consider~\eqref{P_C1_1} and rearrange it to isolate terms that multiply $\tilde{\alpha}_i$ and $\tilde{\beta}_i$ and keeping only the terms that are linear in the measurement noises. After some algebraic manipulation, the average MSE can be expressed as a linear combination of the measurement noise variances $\sigma_{x,i}^2$ and $\sigma_{y,i}^2$ as
\vspace{-3mm}
    \begin{align}\label{equ:MSE_avg_2}
    \hspace{-3mm}\overline{\text{MSE}}\hspace{-0.6mm}\approx\hspace{-0.6mm} \frac{||\mathbf{v}||^2}{\Delta_1^2} \hspace{-1mm}\sum_{i=1}^{M} \hspace{-0.5mm}
    \left( f_x^2 \Delta x_i^2 \mathbb{E}\{\sigma^2_{x,i}\} \hspace{-0.7mm}+\hspace{-0.7mm} f_y^2 \Delta y_i^2 \mathbb{E}\{\sigma_{y,i}^2\} \right),
    \end{align}
where $\mathbf{v} = [B\, C; A\, D; A\, B]$. To compute $\mathbb{E}\{\sigma^2_{x,i}\}$ and $\mathbb{E}\{\sigma^2_{y,i}\}$, we use the following approach. The variance along the $x$-axis can be expressed as $\mathbb{E}\{\sigma_{x,i}^2\} = k_1\mathbb{E}\{1/h_t\} \mathbb{E}\{1/G_{A,S}\} \mathbb{E}\{1/G_{G,A, i}\}$, where $k_1 = \eta \sigma_I^2 A_{cam}/(\alpha\rho_sP_{DL})$, $G_{A,S}$ is the deterministic channel coefficient from the AP to the sensing target, and $G_{S,A,i}$ is the deterministic channel coefficient from the sensing target to the $i$-th pinhole camera. For log-normal turbulence, the reciprocal moment is $\mathbb{E}\{1/h_t\} = \exp(-2\mu_x+2\sigma_x^2)$, which follows directly from a variable transformation on the log-normal pdf. To evaluate $\mathbb{E}\{1/G_{A,S}\}$, we write $G_{A,S} = k_2\text{c}(\theta)^{m_1}$, where $k_2 = (m_1+1)A_p\text{c}\phi/(2\pi d^2)$, and $\text{c}(\theta) = \mathbf{Q}_m \cdot(\mathbf{p}_{S}-\mathbf{p}_{A})/\|\mathbf{p}_{S}-\mathbf{p}_{A}\|$. Expanding $\text{c}(\theta)$ and using the small-angle assumptions $\text{c}\theta_R\approx 1$, $\text{c}\theta_P\approx 1$, $\text{s}\theta_R\approx \theta_R$, and $\text{s}\theta_P\approx \theta_P$, the irradiance-direction cosine at the target can be approximated as \(
    \text{c}(\theta) \approx \alpha_R'\theta_R+\alpha_P'\theta_P+ \bar{C}'\),
where $\bar{C'} = \Delta z'$, $\alpha_R' = \text{s}\theta_D\Delta x'-\text{c}\theta_D\Delta y'$, and $\alpha_P' = \text{c}\theta_D\Delta x'+\text{s}\theta_D\Delta y'$ are geometry-dependent constants. Here $\Delta x' = (x_S-x_A)/d_{S,A}$, $\Delta y' = (y_S-y_A)/d_{S,A}$, $\Delta x' = (z_S-z_A)/d_{S,A}$, and $d_{S,A} = \sqrt{(\Delta x')^2+(\Delta y')^2+(\Delta z')^2}$. Since $\theta_R$, $\theta_Y$, and $\theta_P$ are independent Gaussian random variables, $\text{c}(\theta)$ is also Gaussian distributed with mean $\mu_c = \bar{C}'$ and variance $\sigma_2^2 =\alpha_R'^2\sigma_{\theta_R}^2+ \alpha_P'^2\sigma_{\theta_P}^2$. It follows that 
$\mathbb{E}\{1/G_{A,S}\} = \frac{1}{k_2}\mathbb{E}\{(\text{c}(\theta))^{-m_1}\}$. Using a second order Taylor expansion $\text{c}(\theta)^{-m_1}$ around $\mu_c$, we obtain
\vspace{-2mm}
    \begin{equation}
    \hspace{-2mm}\mathbb{E}\!\left\{\frac{1}{G_{A,S}}\right\}
    \approx
    \frac{1}{k_{2}}
    \left(
    \mu_c^{-m_1}
    + 
    \frac{m_1(m_1+1)}{2}\,
    \mu_c^{-m_1-2}\sigma_c^{2}
    \right).
    \label{eq:EGAS_inverse_m1}
    \end{equation}
Using a similar procedure, the expectation $\mathbb{E}\{\frac{1}{G_{S,A,i}}\}$ can be approximated as
\vspace{-2mm}
    \begin{equation}
    \mathbb{E}\!\left\{\frac{1}{G_{S,A,i}}\right\}
    \approx
    \frac{1}{k_{3,i}}
    \left(
    \frac{1}{\mu_{c,i}}
    +
    \frac{\sigma_{c,i}^{2}}{\mu_{c,i}^{3}}
    \right),
    \label{eq:EGAS_inverse_m2}
    \end{equation}
where $k_{3,i}= (m_1+1)A_p\text{c}^{m_1}(\theta_i)/(2\pi d^2)$, and $\mu_{c,i}$ and $\sigma_{c,i}^{2}$ denote the mean and variance of the direction cosine at the $i$-th camera, respectively. Finally,  
substituting~\eqref{equ:MSE_avg},~\eqref{eq:EGAS_inverse_m1}, and~\eqref{eq:EGAS_inverse_m2} into~\eqref{equ:MSE_avg_2}, we obtain the closed-form approximation for the average MSE given in~\eqref{equ:final_MSE}, shown at the top of the next page. Here $A_1 = Mf_x^2$, $B_1 = Mf_y^2$, $C_1 = f_x^2\sum_{i=1}^M x_{c,i}/z_{c,1}$, and $D_1 = f_y^2\sum_{i=1}^M y_{c,i}/z_{c,1}$.

\begin{figure*}
    \begin{align}\label{equ:final_MSE}
        \overline{\text{MSE}} \hspace{-0.6mm}\approx\hspace{-0.6mm} \frac{((B_1C_1)^2\hspace{-0.6mm}+\hspace{-0.6mm}(A_1D_1)^2\hspace{-0.6mm}+\hspace{-0.6mm}(A_1B_1)^2)k_1e^{(-2\mu_x+2\sigma^2)}}{\Delta_1^2k_{2}\mu_c^{m_1}} \hspace{-1mm}
    \left(1
    \hspace{-0.6mm}+\hspace{-0.6mm} 
    \frac{m_1(m_1\hspace{-0.6mm}+\hspace{-0.6mm}1)}{2\mu_c^{2}}
    \sigma_c^{2}
    \right)\hspace{-1mm}\sum_{i=1}^{M} \hspace{-1mm}\frac{1}{k_{3,i}}\hspace{-0.5mm}
    \left( f_x^2 \Delta x_i^2  \hspace{-0.7mm}+\hspace{-0.7mm} f_y^2 \Delta y_i^2  \right)\hspace{-1mm}
    \left(
    \frac{1}{\mu_{c,i}}
    \hspace{-1mm}+\hspace{-1mm}
    \frac{\sigma_{c,i}^{2}}{\mu_{c,i}^{3}}
    \right),
    \end{align}
    \hrule
    \vspace{-5mm}
\end{figure*}

\vspace{-3mm}
\section{Achievable Rate Analysis}
In this section, the average achievable rate for the uplink communication of the EH sensor is derived. Based on the instantaneous achievable rate bound given in~\eqref{equ:inst1}, approximate closed-form expressions for the average rate are obtained for the considered system setup. By averaging over the channel distributions, the average achievable rate can be expressed as \(R_{\mathrm{UL}}^E= \mathbb{E}_h\left\{R_{\mathrm{UL}}\right\}\), where $\mathbb{E}_h\left\{\cdot\right\}$ denotes expectation over the channel conditions.
Assuming that the oceanic-current-induced turbulence and the channel variations caused by ship attitude dynamics are statistically independent, the average achievable rate can be expressed, with the aid of~\eqref{equ:EH1} and~\eqref{equ:inst1}, as
\vspace{-3mm}
\begin{align}\label{equ:AR1}
        \hspace{-3mm} R_{\mathrm{UL}}^E
        \hspace{-1mm}&=\hspace{-0.5mm}\frac{1\hspace{-0.5mm}
        -\hspace{-0.5mm}\alpha}{2}\hspace{-2mm}\int_{0}^{\infty}\hspace{-2mm}\int_{0}^{\infty}\hspace{-2mm}\int_{0}^{\infty}\hspace{-1mm}\log_{2} \left(
        1 \hspace{-0.5mm}+\hspace{-0.5mm} \kappa_1 h_t^4
        g_{E,A}^2 \ln^2(1\hspace{-0.5mm}+\hspace{-0.5mm}\kappa_2h_tg_{A,E})
        \right)\nonumber \\
        &\times \hspace{-1mm} f_{h_t}(h_t)f_{G_{E,A}}(g_{E,A})f_{G_{A,E}}(g_{A,E})dh_tdg_{E,A} dg_{A,E},\hspace{-1.5mm}
    \end{align}
where $\kappa_1 = ef^2v_t^2\alpha^2 P_{DL}^2/(2\pi(1-\alpha)^2\sigma_n^2)$, $\kappa_2 = P_{DL}/I_{0}$, $G_{A,E}$ denotes the deterministic channel coefficient from the AP to EH sensor, and $G_{E,A}$ represents the deterministic channel coefficient from the EH sensor to AP. $f_{G_{E,A}}(g_{E,A})$, and $f_{G_{A,E}}(g_{A,E})$ denote the pdfs of $G_{E,A}$ and $G_{A,E}$, respectively. 

To evaluate the achievable rate, the pdfs of the channel gains $f_{G_{A,E}}(g_{A,E})$ and $f_{G_{E,A}}(g_{E,A})$ must first be determined. Based on the channel geometry, the irradiance angle $\theta$ in~\eqref{equ:GL} can be expressed as $\text{c}(\theta) = \mathbf{Q}_m \cdot(\mathbf{p}_{E}-\mathbf{p}_{A})/\|\mathbf{p}_{E}-\mathbf{p}_{A}\|$. Using~\eqref{equ:Q_mat}, $\text{c}(\theta)$ can be further expressed with respect to the ship attitude angles, denoted by $\theta_{\text{R}}$, $\theta_{\text{Y}}$, and $\theta_{\text{P}}$, as
\vspace{-2mm}
    \begin{align}
    \text{c}(\theta) &= (\text{c}{\theta_{\text{Y}}}\text{s}{\theta_{\text{P}}}\text{c}{\theta_{\text{R}}}+\text{s}{\theta_{\text{Y}}}\text{s}{\theta_{\text{R}}})\,\Delta x +
    (\text{s}{\theta_{\text{Y}}}\text{s}{\theta_{\text{P}}}\text{c}{\theta_{\text{R}}}- \text{c}{\theta_{\text{Y}}}\text{s}{\theta_{\text{R}}})\,\Delta y \nonumber \\
    &+ \text{c}{\theta_{\text{P}}} \text{c}{\theta_{\text{R}}}\,\Delta z,
    \label{eq:cos_theta}
    \end{align}
where $\Delta x = (x_{E}-x_{A})/d_{E,A}$, $\Delta y = (y_{E}-y_{A})/d_{E,A}$, $\Delta z = (z_{E}-z_{A})/d_{E,A}$, and $d = \sqrt{ (\Delta x)^2 + (\Delta y)^2 + (\Delta z)^2 }$. For small roll and pitch angles $(\theta_R,\theta_P)$, the approximations $\text{c}\theta_R \!\approx\! 1$, $\text{c}\theta_P \!\approx\! 1$, $\text{s}\theta_R \!\approx\! \theta_R$, and $\text{s}\theta_P \!\approx\! \theta_P$ hold. Using these relations, $\text{c}(\theta)$ can be linearized as
\vspace{-2mm}
    \begin{align}
    \text{c}(\theta) \approx \alpha_R\theta_R+\alpha_Y\theta_Y+\alpha_P\theta_P+ \bar{C},
    \label{eq:cos_theta_linear}
    \end{align}
where $\bar{C} = \Delta z$, $\alpha_R = \text{s}\theta_D\Delta x-\text{c}\theta_D\Delta y$, $\alpha_P = \text{c}\theta_D\Delta x+\text{s}\theta_D\Delta y$, and $\alpha_Y=0$ are geometry-dependent constants.

The distribution of $\text{c}(\theta)$ can be approximated using a sine-weighted Gaussian model. This arises from a Gaussian-distributed angle $\theta$ transformed via the cosine function under the small-angle assumption. The resulting pdf is given by
\vspace{-2mm}
    \begin{align}
    f_{\text{c}\theta}(x) \approx 
    \frac{1}{\sqrt{2\pi}\sigma_\text{eff}} 
    \frac{1}{\sqrt{1-x^2}} 
    \exp\!\left[-\frac{(\cos^{-1} x - \bar{\theta})^2}{2\sigma_\text{eff}^2}\right],
    \label{eq:pdf_costheta}
    \end{align}
where $\bar{\theta} = \cos^{-1}(\bar{C})$, and the effective variance is
$\sigma_\text{eff}^2 = \alpha_R^2 \sigma_R^2 + \alpha_P^2 \sigma_P^2 + \alpha_Y^2 \sigma_Y^2$. Given the irradiance-to-channel-gain relationship $G_{A,E} = K\text{c}^{m_1}(\theta)$, the pdf of $G_{A,E}$ can be derived using a standard variable transformation and is approximated as
\vspace{-4mm}
    \begin{align}\label{equ:f_GAE}
    f_{G_{A,E}}(g) &\approx 
    \frac{1}{m_1 K^{1/m_1} g^{1-1/m_1} \sqrt{2\pi}\sigma_\text{eff}} 
    \frac{1}{\sqrt{1 - (g/K)^{2/m_1}}} \nonumber \\
    &\times\exp\!\left[-\frac{(\cos^{-1} ((g/K)^{1/m_1}) - \bar{\theta})^2}{2\sigma_\text{eff}^2}\right],
    \end{align}
where $K = (m_1+1)A_p\text{c}(\phi)/(2\pi d^2)$. Similarly, the pdf of the uplink channel gain $G_{E,A}$ can be approximated as
\vspace{-2mm}
    \begin{align}\label{equ:f_GEA}
    f_{G_{E,A}}(g) &\approx 
    \frac{1}{K_1 \sqrt{2\pi}\sigma_{\text{eff}}} 
    \frac{1}{\sqrt{1 - (g/K_1)^{2}}} \nonumber \\
    &\times\exp\!\left[-\frac{(\cos^{-1} ((g/K_1)) - \bar{\theta})^2}{2\sigma_{\text{eff}}^2}\right],
    \end{align}
where $K_1 = (m_1+1)A_p/(2\pi d^2)$.

\begin{figure*}[ht]
    \begin{align}
    R_{\mathrm{UL}}^E \hspace{-1mm}\approx\hspace{-1mm}
    \frac{1-\alpha}{4\pi^{5/2}\sigma_{\rm eff}^2}
    \hspace{-1mm}\sum_{i=1}^{N_1}\hspace{-0.5mm}\sum_{j=1}^{N_2}\hspace{-0.5mm}\sum_{k=1}^{N_3}
    w_i\,W_j\,\tilde W_k 
    \log_{2}\!\Bigg(1
    \hspace{-1mm}+\hspace{-1mm}\kappa_1\, e^{4X_i}\,K_1^2\text{c}^2\theta_{1,k}\,
    \Big(\ln\!\big(1+\kappa_2\, e^{X_i}\, K\text{c}^m\theta_j\big)\Big)^2
    \Bigg)
    \exp\!\Bigg(\hspace{-1mm}-\hspace{-1mm}\frac{(\theta_j\hspace{-1mm}-\hspace{-1mm}\bar\theta)^2+(\theta_{1,k}\hspace{-1mm}-\hspace{-1mm}\bar\theta_1)^2}{2\sigma_{\rm eff}^2}\Bigg).
    \label{eq:final_rate}
    \end{align}
    \hrule
    \vspace{-5mm}
\end{figure*}

To evaluate the achievable rate, the triple integral over $h_t$, $G_{A,E}$, and $G_{E,A}$ can be decomposed into nested expectations over Gaussian-distributed variables. By applying the variable transformation $X = \ln(h_t)$, the pdf of $h_t$ is converted to a Gaussian distribution. The inner integral can then be efficiently approximated using Gaussian quadrature as
    \begin{align}
    \hspace{-2mm}I_1
    \hspace{-1mm}\approx\hspace{-1mm} \frac{1}{\sqrt{\pi}} 
    \hspace{-1mm}\sum_{i=1}^{N_1} w_i \hspace{-1mm}\,
    \log_{2}\!\left(1\hspace{-1mm}+\hspace{-1mm}\kappa_1 e^{4X_i} g_{E,A}^2 
    \ln^2(1\hspace{-1mm}+\hspace{-1mm}\kappa_2 e^{X_i} g_{A,E})\right),\hspace{-0.5mm}
    \label{eq:I1}
    \end{align}
where the quadrature nodes are $X_i = \mu_X + \sqrt{2}\sigma_X t_i$, with $\mu_X = 2\mu_x$ and $\sigma_X = 2\sigma_x$, and $w_i$, $t_i$ are the corresponding Gaussian weights and points. Next, the integration over $f_{G_{E,A}}(g_{E,A})$ can be performed by substituting $g_{E,A} = K_1\text{c}(\theta_1)$, yielding
    \begin{align}
    I_2
    &\approx \frac{1}{\pi} 
    \sum_{i=1}^{N_1} w_i \sum_{k=1}^{N_2} \tilde{W}_k
    \log_{2}\!\big(1+\kappa_1 e^{4X_i} (K_1\text{c}{\theta_{1,k}})^2 \\
    &\times\ln^2(1+\kappa_2 e^{X_i} g_{A,E})\big)
    f_{G_{E,A}}(K_1\text{c}(\theta_{1,k}))K_1\text{s}(\theta_{1,k}), \nonumber
    \end{align}
where $\theta_{1,k} = \frac{\pi}{4}(x_k+1)$ and $\tilde{W}_k=\frac{\pi}{4}\tilde{w}_k$ correspond to the quadrature nodes and weights. Finally, integrating over $f_{G_{A,E}}(g_{A,E})$ using $g_{A,E} = K\text{c}^{m_1}(\theta)$ gives
    \begin{align}\label{equ:final_rate}
    I_3
    &\approx \frac{1}{\pi^{3/2}} 
    \sum_{i=1}^{N_1} w_i \sum_{k=1}^{N_2} \tilde{W}_k \sum_{j=1}^{N_3} {W}_j
    \log_{2}\big(1+\kappa_1 e^{4X_i} (K_1\text{c}{\theta_{1,k}})^2 \nonumber \\
    &\times \ln^2(1+\kappa_2 e^{X_i} (K\text{c}^{m_1}{\theta_j}))\big)f_{G_{E,A}}(K_1\text{c}(\theta_{1,k}))K_1\text{s}(\theta_{1,k})\nonumber \\
    &\times
    f_{G_{A,E}}(K\text{c}^{m_1}(\theta_j))Km_1\text{c}^{m_1-1}(\theta_j)\text{s}(\theta_j),
    \end{align}
where $\theta_{j} = \frac{\pi}{4}(y_j+1)$ and ${W}_j=\frac{\pi}{4}\tilde{w}_j$ are the quadrature nodes and weights for the integration over $G_{A,E}$. Using~\eqref{equ:f_GAE},~\eqref{equ:f_GEA}, and~\eqref{equ:final_rate}, the achievable rate upper bound can be approximated as
in~\eqref{eq:final_rate} at the top of the next page, which provides a tractable approximation for the achievable rate under the considered random orientation and geometric channel effects.

\vspace{-3mm}
\section{Numerical Results and Discussion}
This section presents numerical results that validate the analytical derivations, demonstrate the communication–sensing trade-off, and highlight the influence of key system and channel parameters. The simulation setup considers an AP, an EH sensor, and a sensing target located at coordinates $(0, 0, 10)$, $(-2, -2, 0)$, and $(2, 2, 0)$, respectively. Unless otherwise specified, the parameters used in all simulations are $\rho = 1.33$, $c = 0.1$, $A_{p} = 10^{-3}$~m$^2$, $r = 0.5$~A/W, $\Phi_{\text{FoV}} = 90^{\circ}$, $\mu_x = -0.1$, $\sigma_x^2 = 0.1$, $A_{\text{cam}} = 10^{-3}$~m$^2$, $f_x = 0.05$~m, $f_y = 0.05$~m, $\sigma_{\theta_R}^2 = 10$, $\sigma_{\theta_P}^2 = 10$, $\sigma_{\theta_Y}^2 = 10$, $\sigma_I^2 = 10^{-6}$, $f = 0.9$, $v_t = 25$~mV, $r_{\text{PV}} = 0.9$~A/W, and $I_0 = 10^{-9}$~A.

\begin{figure}[!t]
    \centering
    \includegraphics[width=0.7\linewidth]{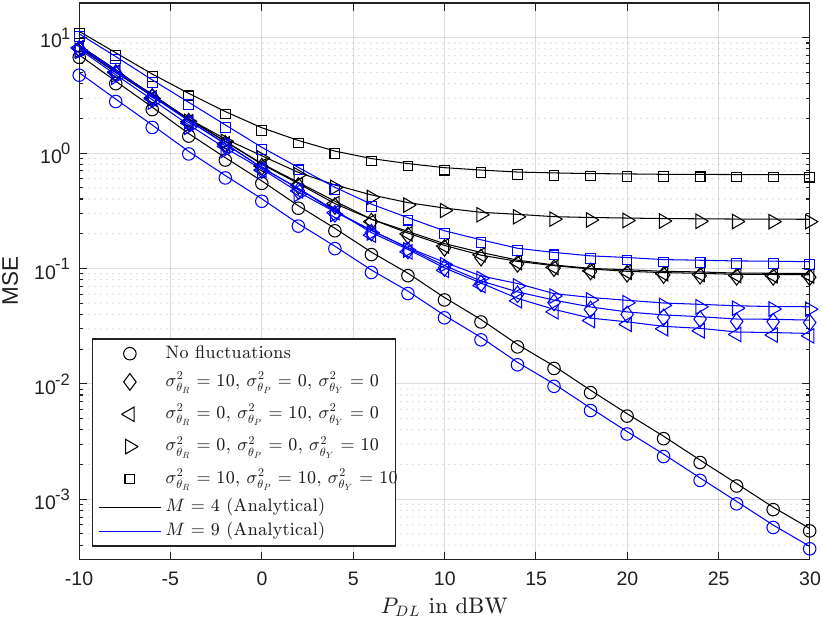}
    \vspace{-3mm}
    \caption{Average sensing MSE versus downlink transmit power $P_{DL}$.}
    \label{fig:Result1}
    \vspace{-3mm}
\end{figure}

Fig.~\ref{fig:Result1} depicts the average sensing MSE as a function of the downlink transmit power, $P_{DL}$. Analytical and simulation results show excellent agreement, confirming the accuracy of the derived expressions. The MSE performance is shown for various attitude conditions and numbers of pinhole cameras. As expected, the MSE decreases with increasing $P_{DL}$ due to the improved signal-to-noise ratio. However, at high attitude variances, an error floor emerges because of degraded channel conditions between the AP and the sensing target. Increasing the number of pinhole cameras significantly enhances MSE performance and mitigates the error floor, particularly under severe attitude fluctuations, due to high 
spatial diversity. For instance, a 10~dB improvement in MSE is observed for $M = 9$ compared to $M = 4$ when $\sigma_{\theta_R}^2 = \sigma_{\theta_P}^2 = \sigma_{\theta_Y}^2 = 10$.

\begin{figure}[!t]
    \centering
    \includegraphics[width=0.7\linewidth]{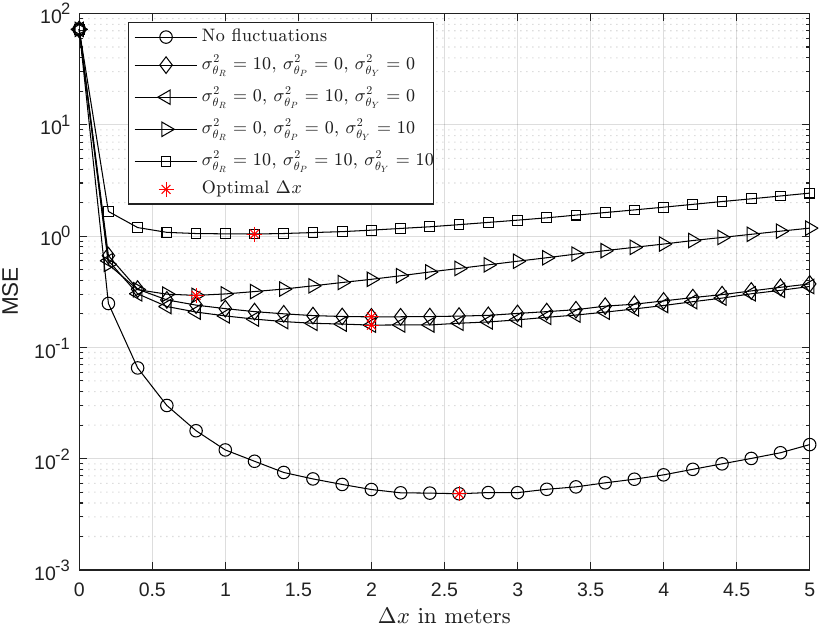}
    \vspace{-3mm}
    \caption{Average sensing MSE versus distance between pinhole cameras.}
    \label{fig:Result2}
    \vspace{-6mm}
\end{figure}

Fig.~\ref{fig:Result2} illustrates the average sensing MSE as a function of the distance between pinhole cameras, $\rho_x/\rho_y$, under different attitude variation conditions. When the pinhole separation is small, higher channel gains are achieved. However, increasing the separation enhances spatial diversity. This trade-off leads to an optimal camera spacing, which depends on the level of attitude variation. Specifically, the optimal $\rho_x$ decreases as the attitude variation increases. For example, the optimal $\rho_x$ values are approximately $2.6$ and $1.2$ for the no-fluctuation and $\sigma_{\theta_R}^2 = \sigma_{\theta_P}^2 = \sigma_{\theta_Y}^2 = 10^{\circ}$ cases, respectively, as the degradation in average channel gain becomes more dominant under severe attitude variations.

\begin{figure}[!t]
    \centering
    \includegraphics[width=0.73\linewidth]{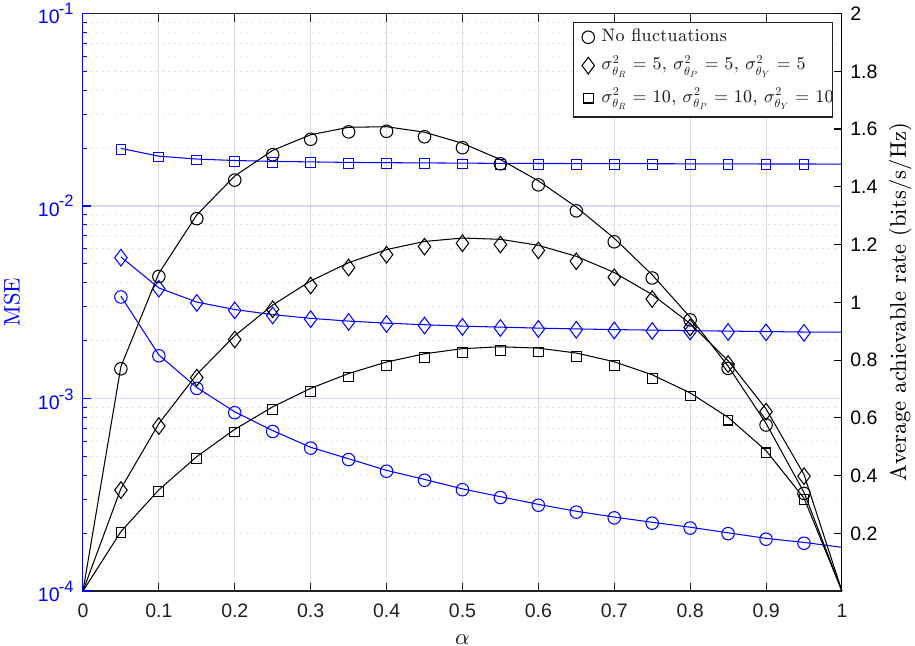}
    \vspace{-3mm}
    \caption{Average sensing MSE and achievable rate versus time allocation parameter $\alpha$.}
    \label{fig:Result3}
    \vspace{-6mm}
\end{figure}

Fig.~\ref{fig:Result3} presents the variation of the average sensing MSE and the achievable communication rate with respect to the time allocation parameter $\alpha$. The achievable rate exhibits a clear maximum due to the nature of the harvest-use protocol. The optimal $\alpha$ increases with attitude fluctuation, since higher attitude variance leads to weaker downlink channels that require longer energy harvesting durations. Conversely, the sensing MSE decreases gradually with $\alpha$, as longer exposure times at the cameras reduce sensing noise. This observation clearly illustrates the inherent communication–sensing trade-off in the proposed system. Furthermore, these results emphasize the necessity of jointly optimizing sensing and communication parameters, as optimizing for achievable rate under high attitude fluctuations also benefits sensing performance.

\vspace{-4mm}
\section{Conclusion}
In this paper, we proposed an LPT-enabled underwater O-ISAC system, where a ship/buoy-mounted AP transmits lightwave signals in the downlink to two seabed nodes, namely, ($i$) an EH sensor that harvests optical energy and utilizes it for uplink communication, and ($ii$) a sensing target that reflects light, enabling the AP to estimate its location via an array of pinhole cameras. The ship’s attitude variation was modeled through stochastic roll, pitch, and yaw angles, each following Gaussian distributions under low-to-moderate sea states. Closed-form analytical expressions were derived for the sensing MSE, and the achievable uplink data rate. Analytical and simulation results demonstrated excellent agreement, validating the proposed analytical models. Furthermore, the results provided valuable design insights, including optimal time allocation between downlink and uplink phases, and optimal pinhole camera placement. Finally, the fundamental communication–sensing tradeoff of the proposed O-ISAC system was characterized, offering useful guidelines for future LPT-enabled underwater optical network design.

\vspace{-4mm}
\bibliographystyle{IEEEtran}
\bibliography{IEEEabrv,main_after_CP_comments_v1.00}

\begin{thebibliography}{10}
\providecommand{\url}[1]{#1}
\csname url@samestyle\endcsname
\providecommand{\newblock}{\relax}
\providecommand{\bibinfo}[2]{#2}
\providecommand{\BIBentrySTDinterwordspacing}{\spaceskip=0pt\relax}
\providecommand{\BIBentryALTinterwordstretchfactor}{4}
\providecommand{\BIBentryALTinterwordspacing}{\spaceskip=\fontdimen2\font plus
\BIBentryALTinterwordstretchfactor\fontdimen3\font minus \fontdimen4\font\relax}
\providecommand{\BIBforeignlanguage}[2]{{%
\expandafter\ifx\csname l@#1\endcsname\relax
\typeout{** WARNING: IEEEtran.bst: No hyphenation pattern has been}%
\typeout{** loaded for the language `#1'. Using the pattern for}%
\typeout{** the default language instead.}%
\else
\language=\csname l@#1\endcsname
\fi
#2}}
\providecommand{\BIBdecl}{\relax}
\BIBdecl

\bibitem{Zeng_2017}
Z.~Zeng, S.~Fu, H.~Zhang, Y.~Dong, and J.~Cheng, ``A survey of underwater optical wireless communications,'' \emph{{IEEE} Commun. Surv. Tutor.}, vol.~19, pp. 204--238, 1st Quat. 2017.

\bibitem{Celik_2020}
A.~Celik, N.~Saeed, B.~Shihada, T.~Y. Al-Naffouri, and M.-S. Alouini, ``End-to-end performance analysis of underwater optical wireless relaying and routing techniques under location uncertainty,'' \emph{{IEEE} Trans. Wirel. Commun.}, vol.~19, pp. 1167--1181, Feb. 2020.

\bibitem{Kapila_2021}
K.~W.~S. Palitharathna, H.~A. Suraweera, R.~I. Godaliyadda, V.~R. Herath, and J.~S. Thompson, ``Average rate analysis of cooperative {NOMA} aided underwater optical wireless systems,'' \emph{{IEEE} Open J. Commun. Soc.}, vol.~2, pp. 2292--2310, Sept. 2021.

\bibitem{Xiang_2024}
X.~Yi \emph{et~al.}, ``Aperture-averaged angle-of-arrival fluctuations in oceanic turbulence of arbitrary strength,'' \emph{{IEEE} Trans. Antennas Propag.}, vol.~72, pp. 2631--2642, Mar. 2024.

\bibitem{Newman_1977}
J.~N. Newman, \emph{Marine Hydrodynamics}.\hskip 1em plus 0.5em minus 0.4em\relax Cambridge, MA, USA: The {MIT} Press, 1977.

\bibitem{Kapila_2022}
K.~W.~S. Palitharathna, H.~A. Suraweera, R.~I. Godaliyadda, V.~R. Herath, and Z.~Ding, ``Lightwave power transfer in full-duplex {NOMA} underwater optical wireless communication systems,'' \emph{{IEEE} Commun. Lett.}, vol.~26, pp. 622--626, Mar. 2022.

\bibitem{Panagiotis_2018}
P.~D. Diamantoulakis, G.~K. Karagiannidis, and Z.~Ding, ``Simultaneous lightwave information and power transfer ({SLIPT}),'' \emph{{IEEE} Trans. Green Commun. Netw.}, vol.~2, pp. 764--773, Mar. 2018.

\bibitem{Shuai_2019}
S.~Ma \emph{et~al.}, ``Simultaneous lightwave information and power transfer in visible light communication systems,'' \emph{{IEEE} Trans. Wirel. Commun.}, vol.~18, pp. 5818--5830, Dec. 2019.

\bibitem{Uysal_2021}
M.~Uysal \emph{et~al.}, ``{SLIPT} for underwater visible light communications: Performance analysis and optimization,'' \emph{{IEEE} Trans. Wirel. Commun.}, vol.~20, pp. 6715--6728, Oct. 2021.

\bibitem{Wen_2024}
Y.~Wen, F.~Yang, J.~Song, and Z.~Han, ``Optical integrated sensing and communication: Architectures, potentials and challenges,'' \emph{{IEEE} Internet Things Mag.}, vol.~7, pp. 68--74, July 2024.

\bibitem{Alizera_2025}
\BIBentryALTinterwordspacing
A.~G. Khorasgani, M.~Mirmohseni, and A.~Elzanaty, ``Optical {ISAC}: Fundamental performance limits and transceiver design,'' \emph{arXiv preprint arXiv:2408.11792}, 2025. [Online]. Available: \url{https://arxiv.org/abs/2408.11792}
\BIBentrySTDinterwordspacing

\bibitem{Zhang_2025}
R.~Zhang, Y.~Shao, M.~Li, L.~Lu, and Y.~C. Eldar, ``Optical integrated sensing and communication with light-emitting diode,'' \emph{{IEEE} Internet Things J.}, vol.~12, pp. 12\,896--12\,911, May. 2025.

\bibitem{Lapidoth_2009}
A.~Lapidoth, S.~M. Moser, and M.~A. Wigger, ``On the capacity of free-space optical intensity channels,'' \emph{{IEEE} Trans. Infor. Theory}, vol.~55, pp. 4449--4461, Oct. 2009.

\end{thebibliography}

\end{document}